\documentclass[utf8]{FrontiersinHarvard} 

\usepackage{url,hyperref,lineno,microtype,subcaption}
\usepackage[onehalfspacing]{setspace}
\usepackage{threeparttable}

\newcommand\arcdeg{\mbox{$^\circ$}}
\newcommand\arcmin{\mbox{$^\prime$}}
\newcommand\arcsec{\mbox{$^{\prime\prime}$}}
\newcommand\farcs{\mbox{$.\!\!^{\prime\prime}$}}

\newcommand{\apjl}{\textit{ApJL}}
\newcommand{\apjs}{\textit{ApJS}}
\newcommand{\aap}{\textit{A\&A}}

\def\keyFont{\fontsize{8}{11}\helveticabold }
\def\firstAuthorLast{Shi {et~al.}} 
\def\Authors{DongDong Shi\,$^{1,*}$, XianZhong Zheng\,$^{2,*}$, Zhizheng Pan\,$^{3}$, Yu Luo\,$^{4}$, Hongxia Deng\,$^{1}$, Qunzhi Hua\,$^{1}$, Xinyu Luo\,$^{1}$ and Qiming Wu\,$^{1}$}
\def\Address{$^{1}$Center for Fundamental Physics, School of Mechanics and Optoelectronic Physics, Anhui University of Science and Technology, Huainan 232001, China \\
$^{2}$Tsung-Dao Lee Institute and Key Laboratory for Particle Physics, Astrophysics and Cosmology, Ministry of Education, Shanghai Jiao Tong University, Shanghai, 201210, China \\
 $^{3}$ Purple Mountain Observatory, Chinese Academy of Sciences, 10 Yuan Hua Road, Nanjing, Jiangsu 210023, China \\
$^{4}$ Department of Physics, School of Physics and Electronics, Hunan Normal University, Changsha 410081, China} 
\def\corrAuthor{DongDong Shi,XianZhong Zheng}

\def\corrEmail{ddshi@aust.edu.cn,xzzheng@sjtu.edu.cn}

\begin{document}
\onecolumn
\firstpage{1}


\title[Diffuse galaxies in COSMOS]{Searching for Nearby Diffuse Dwarf Galaxies in the COSMOS Field} 

\author[\firstAuthorLast ]{\Authors} 
\address{} 
\correspondance{} 

\extraAuth{}

\maketitle

\begin{abstract}

\section{}
It remains challenging to systematically survey nearby diffuse dwarf galaxies and address the formation mechanism of this population distinguishing from regular ones.  We carry out a pilot search for these galaxies in the COSMOS field using the deep \textit{HST}/F814W imaging data. 
We report three diffuse dwarf galaxies satisfying the criteria: (1) redshift $z<0.2$, (2) effective radius $r_{\rm e}>1\farcs0$, and (3) central surface brightness $\mu_{\rm 0}>24$\,mag\,arcsec$^{-2}$. Two of the three galaxies, COSMOS-UDG1 and COSMOS-UDG2, are recognized as ultra-diffuse galaxies (UDGs) with redshift $z=0.130$ and $0.049$, respectively. The third galaxy, COSMOS-dw1, is spectroscopically confirmed as a dwarf galaxy at $z=0.004$. 
We derive the physical properties through fitting their spectral energy distributions (SEDs) extracted from deep multiwavelength observations. COSMOS-dw1 has a stellar mass of $5.6_{-2.7}^{+2.5}\times10^{6}$\,M$_{\odot}$, harboring neutral hydrogen gas of mass $4.90\pm0.90\times10^{6}$\,M$_{\odot}$, hinting that this galaxy may be in the nascent stages of quenching. The estimated dynamical mass of $3.4\times10^{7}\,M_{\odot}$ further suggests that COSMOS-dw1 is predominantly of dark matter. COSMOS-UDG1 and COSMOS-UDG2 exhibit comparable stellar masses of $\sim 2\times10^{8}$\,M$_{\odot}$.  Notably, COSMOS-UDG1 is younger and more metal-rich than COSMOS-UDG2 and COSMOS-dw1. Conversely, COSMOS-UDG2 and COSMOS-dw1 have similar stellar metallicities, yet COSMOS-UDG2 is older than COSMOS-dw1.   
All three galaxies adhere to the stellar mass-metallicity relation (MZR) for dwarf galaxies in the local Universe, implying they belong to the dwarf galaxy population.

\tiny
 \keyFont{ \section{Keywords:} Galaxy formation; Galaxy evolution; COSMOS field; Ultra-diffuse galaxies; Dwarf galaxies; Extragalactic astronomy} 
\end{abstract}

\section{Introduction}

The extremely low surface brightness (LSB) galaxies in the Universe provide crucial insights into galaxy formation and evolution, particularly concerning the role of dark matter and the mechanisms shaping diverse galaxy morphologies \citep[e.g.,][]{Impey1997,Bullock2017}. Ultra-diffuse galaxies (UDGs) are distinguished by their extremely low central surface brightness $\mu(g,0) > 24$\,mag\,arcsec$^{-2}$, large half-light radius ($r_{\rm e} > 1.5$\,kpc) comparable to that of typical $L^\ast$ galaxies and relatively low stellar mass ($<\sim 10^{8}$\,$M_{\odot}$), which is two orders of magnitude smaller than that of $L^\ast$ galaxies \citep{vanDokkum2015a}. It remains unclear whether these extreme properties arise from distinct formation processes compared to normal galaxies.

UDGs have been found in a variety of environments, including galaxy clusters \citep[e.g.,][]{vanDokkum2015a, Koda2015, Mihos2015, Munoz2015, Martinez-Delgado2016, vanderBurg2016, Roman2017a, Janssens2017, Lee2017, Lee2020, Iodice2020, Wittmann2017, Janssens2019,Gannon2022,Venhola2022, LaMarca2022a, LaMarca2022b}, groups \citep[e.g.,][]{Smith2016, Merritt2016, Ordenes-Briceno2016, Trujillo2017, Roman2017b, Shi2017, vanderBurg2017, Muller2018, Greco2018b, Somalwar2020, Zaritsky2023, Jones2023} and the fields \citep[e.g.,][]{Bellazzini2017, Leisman2017, Bennet2018, Prole2019, Barbosa2020, Fielder2024, Montes2024}, as well as within cosmic void \citep{Roman2019} and associated with the large-scale structures \citep[e.g.,][]{Roman2017a, Shi2017}. Accumulating observational evidence reveals that UDGs exhibit a diverse range of  properties. They can be red and blue in color \citep[e.g.,][]{Koda2015, Roman2017a, Roman2017b, Shi2017, Leisman2017}, gas-poor and gas-rich in  H\,{\small I} gas content \citep[e.g.,][]{Trujillo2017, Kadowaki2017, Bellazzini2017, Papastergis2017, Spekkens2018, Karunakaran2020,Karunakaran2024}, and have prolate and oblate in geometry/intrinsic ellipticity distribution \citep{Burkert2017}.  Some UDGs contain a high fraction of globular clusters (GCs) \citep[e.g.,][]{Beasley2016a, Beasley2016b, Peng2016, vanDokkum2016, vanDokkum2017, vanDokkum2018b, Toloba2018, Lim2018, Marleau2024, Forbes2025}, and many host a compact nucleus in their central regions \citep[e.g.,][]{Yagi2016, Janssens2017,Lambert2024,Khim2024}. These observations suggest that the UDG population, selected based on observational criteria, may be composed of different populations formed through different pathways. 
For instance, UDGs Dragonfly~44 and Dragonfly~X1 are likely overwhelmingly dominated by dark matter and are considered as ``failed'' galaxies with massive halo ($>5 \times 10^{11}$\,$M_{\odot}$) \citep{vanDokkum2015a, vanDokkum2016, vanDokkum2017}, and UDGs VCC~1287 and Dragonfly~17 are seen as ``failed'' Large Magellanic Cloud (LMC) or M33 with low halo mass ($< 10^{11}$\,$M_{\odot} $) \citep{Beasley2016b, Peng2016, Amorisco2018}. In contrast,  several UDGs have been reported to be tidally disrupted dwarf galaxies, also known as tidal debris or disturbed UDGs \citep{Mihos2015, Mihos2017, Merritt2016, Greco2018a, Fielder2024}. These typically exhibit lower dark matter halos compared to what is expected for their stellar mass \citep{vanDokkum2018a, Toloba2018, Ogiya2018}. The UDGs, NGC1052-DF2 and DF4, contain little or no dark matter  \citep{vanDokkum2018a, vanDokkum2018b, vanDokkum2019, Shen2021, vanDokkum2022}, although the controversy still remains \citep{Trujillo2019, Montes2020}.

Multiple formation scenarios have been proposed to account for the extended nature of UDGs. Galaxy collisions in dense environments are suggested as a mechanism for forming UDGs \citep{Baushev2018}, often leading to prolate rather than oblate morphologies \citep{Burkert2017}. \cite{Carleton2019} proposed that tidal stripping and heating are primary drivers of UDG formation in the dense environments. In less-dense environments, such as poor clusters and galaxy groups, the interaction of the interstellar medium (ISM) with the intra-cluster medium (ICM) is believed to play a pivotal role in shaping UDGs \citep{Levy2007}. 
On the other hand,  \cite{Amorisco2016} contended that UDGs are predominantly dwarf galaxies with extremely high spins. \cite{Leisman2017} and \cite{Spekkens2018} reported that gas-rich UDGs tend to reside in halos of high angular momentum traced by  H\,{\small I} line width,  supporting the high-spin scenario \citep{Amorisco2016, Rong2017}. Furthermore, gas outflow driven by strong feedback from supernovae and massive star winds  in a star-forming galaxy is suggested to cause the expansion of dark matter and stellar disk, ultimately reshaping the galaxy into a faint and extended form \citep{DiCintio2017, Chan2018}.  In addition, \cite{Sales2020} suggested that UDG population is a mixture of normal LSB galaxies typically found in the low-density environments, along with a distinct population whose expansive size and LSB are a result of the impact of cluster tides \citep[e.g.,][]{Tremmel2020}.

\begin{table*}[ht]
\centering
\caption{ The properties of  COSMOS-dw1, COSMOS-UDG1 and COSMOS-UDG2. The parameters listed from top to bottom rows, refer to coordinates, redshift, magnitude in F606W/F814W, color, central surface brightness in F606W/F814W, effective radius in angular and physical size, S{\'e}rsic index\,(n), axis ratio\,(b/a),  stellar mass log(M$_{*}$/M$_{\odot}$), SFR ($M_{\odot}$\,yr$^{-1}$), stellar age (Gyr), log($\tau$/Gyr), stellar metallicity log(Z/Z$_{\odot}$) and H$\small \rm I$ mass, respectively. Note that we correct the extinction, K-correction and cosmological dimming effects.}  \label{tab:tab1}
\begin{threeparttable}
\begin{tabular}{lccc}
\hline 
\hline
 & COSMOS-dw1 & COSMOS-UDG1 & COSMOS-UDG2  \\
\hline
R.A.\,(J2000.0) & 10:00:30.069 & 10:00:37.859 & 10:00:23.752  \\
Decl.\,(J2000.0)& +02:08:59.07 & +02:24:31.86 & +02:22:05.87  \\
redshift & 0.004 & 0.130 & 0.049  \\
$V_{606}$\,(mag) & 19.90$\pm$0.03 & 22.81$\pm$0.11 & 21.68$\pm$0.06 \\
$I_{814}$\,(mag) & 19.79$\pm$0.04 & 22.56$\pm$0.13 & 21.55$\pm$0.07 \\
$V_{606}$-$I_{814}$\,(mag) & 0.11$\pm$0.05 & 0.25$\pm$0.17 & 0.13$\pm$0.09 \\
$\mu(\rm V_{606},0)$\,(mag\,arcsec$^{-2}$) & 24.37$\pm$0.16 & 24.86$\pm$0.11 & 24.48$\pm$0.15 \\
$\mu(\rm I_{814},0)$\,(mag\,arcsec$^{-2}$) & 24.36$\pm$0.19 & 24.75$\pm$0.13 & 24.31$\pm$0.12 \\
r$_{\rm e,I_{814}}$\,($\arcsec$) & 3.42$\pm$0.07 & 1.14$\pm$0.07 & 1.77$\pm$0.08  \\
r$_{\rm e,I_{814}}$\,($\rm kpc$) & 0.29$\pm$0.01 & 2.64$\pm$0.16 & 1.70$\pm$0.08  \\
S{\'e}rsic index\,(n) & 0.20$\pm$0.02 & 0.19$\pm$0.07 & 0.59$\pm$0.08  \\
axis ratio\,(b/a) & 0.65$\pm$0.01 & 0.81$\pm$0.05 & 0.61$\pm$0.08    \\
log(M$_{*}$/M$_{\odot}$) & $6.75_{-0.21}^{+0.19}$ & $8.39_{-0.61}^{+0.33}$ & $8.34_{-0.21}^{+0.17}$ \\
SFR ($M_{\odot}$\,yr$^{-1}$)  & $\sim0.001$ & $\sim0.239$ & $\sim0.008$ \\
age (Gyr) & $4.10_{-2.5}^{+5.24}$ & $1.70_{-1.43}^{+4.41}$ & $5.58_{-3.12}^{+5.10}$ \\
log($\tau$/Gyr) & $0.23_{-0.80}^{+0.60}$ & $0.50_{-0.85}^{+0.97}$ & $0.07_{-0.71}^{+0.52}$ \\
log(Z/Z$_{\odot}$) & $-1.47_{-0.39}^{+0.63}$ & $-0.74_{-0.86}^{+0.54}$ & $-1.44_{-0.40}^{+0.65}$ \\
log(M$_{\rm HI}$/M$_{\odot}$) & $6.69\pm0.08$ & ... & ... \\
\hline
\end{tabular}
\end{threeparttable}
\end{table*}

\begin{figure*}
\begin{center}
\includegraphics[trim=18mm 18mm 18mm 4.4mm,clip,height=0.35\textwidth]{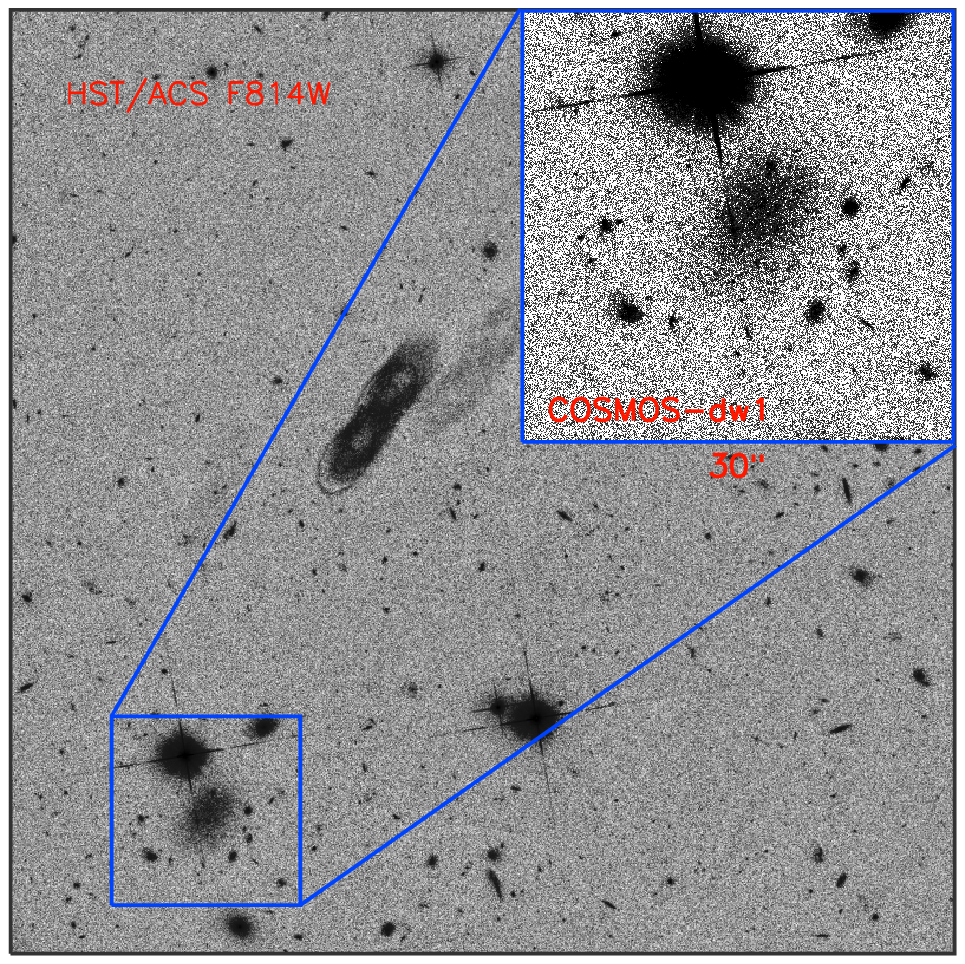}
\includegraphics[trim=18mm 18mm 18mm 4.4mm,clip,height=0.35\textwidth]{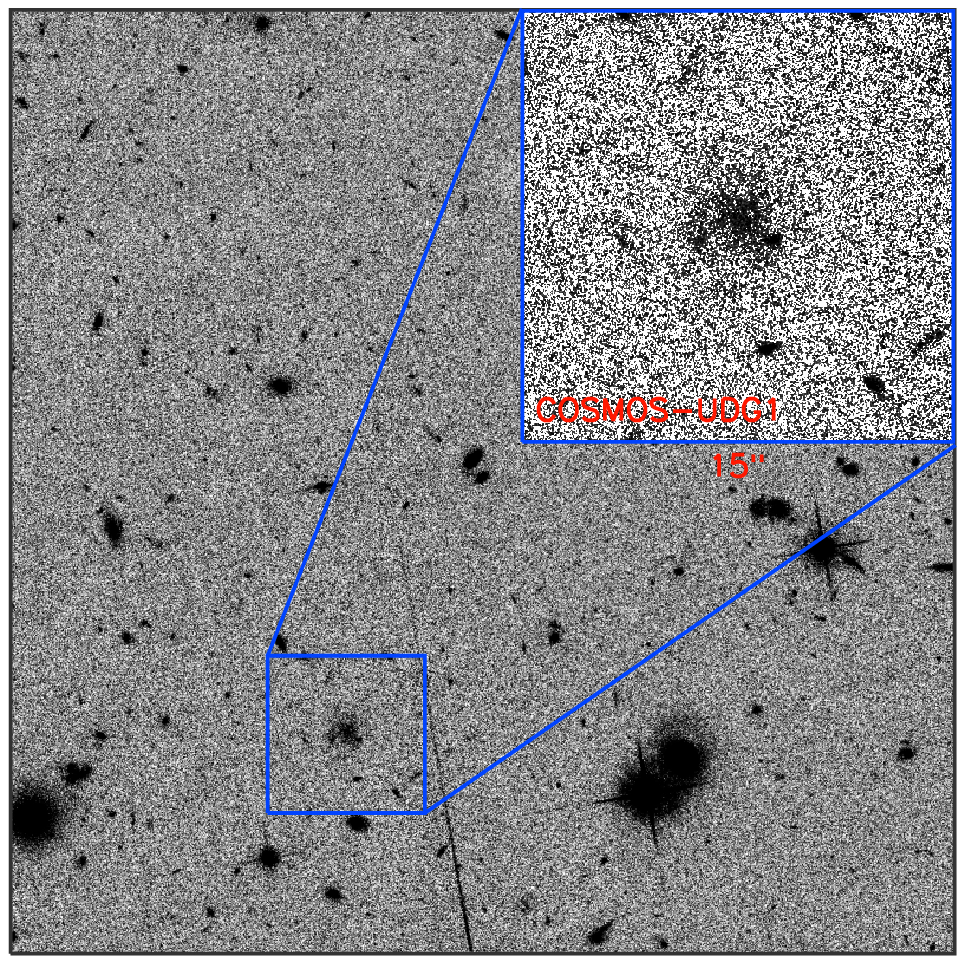}
\includegraphics[trim=18mm 18mm 18mm 4.4mm,clip,height=0.35\textwidth]{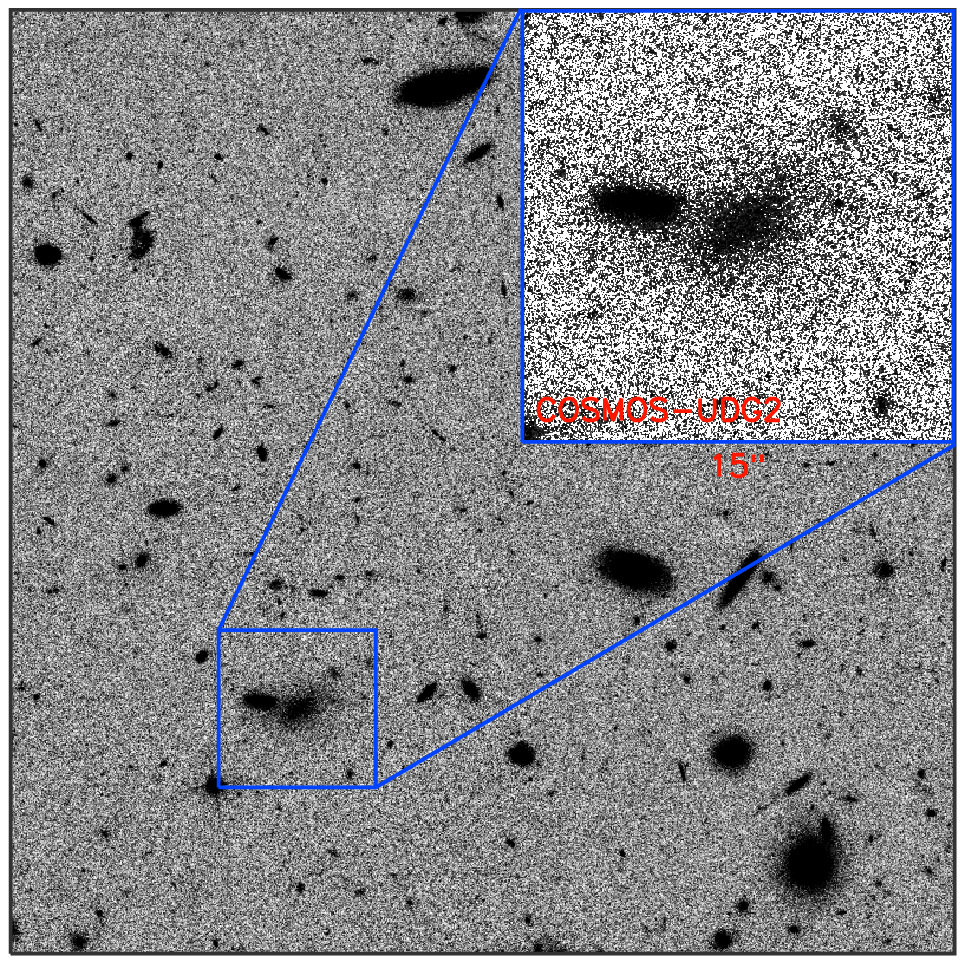}
\caption{The {\it HST}/ACS F814W stamps of three COSMOS diffuse galaxies:  COSMOS-dw1 ({\it left}), COSMOS-UDG1 ({\it middle}),  and COSMOS-UDG2 ({\it right}). The size of the stamps is $90\arcsec \times 90\arcsec$. The inner box shows a region of 30$\arcsec \times 30\arcsec$ for COSMOS-dw1, and 15$\arcsec \times 15\arcsec$ for COSMOS-UDG1 and COSMOS-UDG2.}
\label{fig:fig1}
\end{center}
\end{figure*}

More observational efforts are eagerly demanded  to determine the physical properties of UDGs for a better understanding of their origin. Notably, spectroscopic observations are crucial for revealing the properties of stellar populations, metallicity and kinematics.  However, it is very expensive to obtain good-quality spectroscopic data for UDGs even with 10 meter-class telescopes.  To date, only $\leq$100 UDGs have been spectroscopically observed , and most of them are cluster UDGs \citep[e.g.,][]{vanDokkum2015b, vanDokkum2016, Martinez-Delgado2016, Trujillo2017, Kadowaki2017, Gu2018, Ferre-Mateu2018, Ruiz-Lara2018, Buzzo2022, Buzzo2024a, Gannon2024, Shen2024}. These cluster UDGs are mainly dominated by old and metal-poor populations \citep{Kadowaki2017, Ferre-Mateu2018, Ruiz-Lara2018, Iodice2023}.  On the other hand, some UDGs (e.g., DGSAT\,1 and UGC\,2162) in low-density environments seem to consist of relatively young and high-metallicity stellar populations \citep{Martinez-Delgado2016, Trujillo2017, Pandya2018}. Moreover, some blue UDGs appear to be gas-rich galaxies so that the 21\,cm line can be used to measure the distance of UDGs and test their formation mechanisms \citep{Trujillo2017, Papastergis2017, Bellazzini2017, Leisman2017, Shi2017, Spekkens2018}. Additionally, only a few UDGs has been observed for measuring their stellar kinematics through spectroscopy \citep[e.g.,][]{Chilingarian2019,vanDokkum2019,Iodice2023}, revealing that their dark matter content and velocity profile are diverse \citep[e.g.,][]{Emsellem2019, Kravtsov2024}. Most UDGs have a large dark matter fraction than dwarf galaxies with similar luminosities, but several UDGs contain little or no dark matter \citep{vanDokkum2018a, vanDokkum2018b, vanDokkum2019, Shen2021, vanDokkum2022}.  \cite{Buttitta2025} mapped the stellar kinematics of some UDGs in the Hydra-I cluster, finding that  seven UDGs are in a mild rotation and five UDGs show no evidence of rotation.  Recently, some works explored the stellar populations of UDGs using the multiwavelength SED fitting, and concluded that their properties are diverse \citep[e.g.,][]{Pandya2018,Gu2018,Buzzo2022,Buzzo2024a}. Therefore, further investigation of the physical properties of these diffuse galaxies is imperative.

In this work, we systematically search for extremely LSB galaxies in the COSMOS field. The availability of pre-existing deep multiwavelength observations, spanning from the ultraviolet (UV) to the radio, enables us to delve into the properties of these galaxies in detail. In Section~\ref{sec:observe}, we describe the selection of the three diffuse galaxies and the photometric data. Section~\ref{sec:analysis} presents the photometry and analysis. finally, we discuss and summarize our results in Section~\ref{sec:discussion}. We adopt a  cosmology with $\Omega_{\rm M} = 0.3$, $\Omega_{\rm \Lambda} =0.7$ and $ H\rm_0 = 70$\,km\,s$^{-1}$\,Mpc$^{-1}$, and the AB magnitude system throughout this work.

\section{Target Selection and Data} \label{sec:observe}

We carry out a search for UDGs within the central $36\arcmin \times 14\arcmin$ region of the COSMOS field, where $\it{HST}$/ACS F606W ($V_{606}$) and F814W ($I_{814}$) observations are available from the 3D-HST/CANDELS survey \citep{vanDokkum2013, Momcheva2016}. We made use of the 3D-HST redshift and photometric catalog of 33\,879 objects (the v4.1.5 release) based on the $I_{814}$ detection to select UDG candidates. We limit redshift at $z<0.2$, and apply the selection criteria  of $\mu(I_{814},0)>24.0$\,mag\,arcsec$^{-2}$ and effective radius $r_{\rm e}>1.5$\,kpc to the catalog, yielding a sample of $\sim$20 objects as the UDG candidates. We visually examine the $I_{814}$ images of these targets to get rid of false sources (e.g., blending and compact sources, contamination light from the outer of saturate stars). Finally, we obtained three diffuse galaxies, named as COSMOS-dw1, COSMOS-UDG1 and COSMOS-UDG2. Their $I_{814}$ images are shown in Figure~\ref{fig:fig1}.  Of them, COSMOS-dw1 is not included in the 3D-HST catalog. It was found when visually checked the $I_{814}$ images. 
We cross correlated these objects with the COSMOS2015 catalog \citep{Laigle2016}, finding that the three objects are all included. We note that COSMOS-dw1 has been confirmed by LRIS on Keck~I, and the spectroscopic redshift (spec-$z$) is 0.0041 \citep{Polzin2021}. The spec-$z$ of COSMOS-dw1 in radio observations with Five-hundred-meter Aperture Spherical radio Telescope (FAST) and MeerKAT GHz Tiered Extragalactic Explorations (MIGHTEE) H$\small \rm I$ survey is 0.004 \citep{Pan2024,Heywood2024}. No spec-$z$ is available for the rest two objects (although COSMOS-UDG1 was observed through the Very Large Telescope (VLT)/VIMOS spectrograph, there is still no spec-$z$ due to the low Signal-to-Noise Ratio (SNR) spectrums \citep{Lilly2007}). The COSMOS2015 catalog provides photo-$z$ of the three objects as  0.005$\pm$0.0034, 0.158$\pm$0.008, 0.044$\pm$0.011, respectively.   

\begin{figure*}
\begin{center}
\includegraphics[trim=0mm 0mm 0mm 0mm,clip,height=0.43\textwidth]{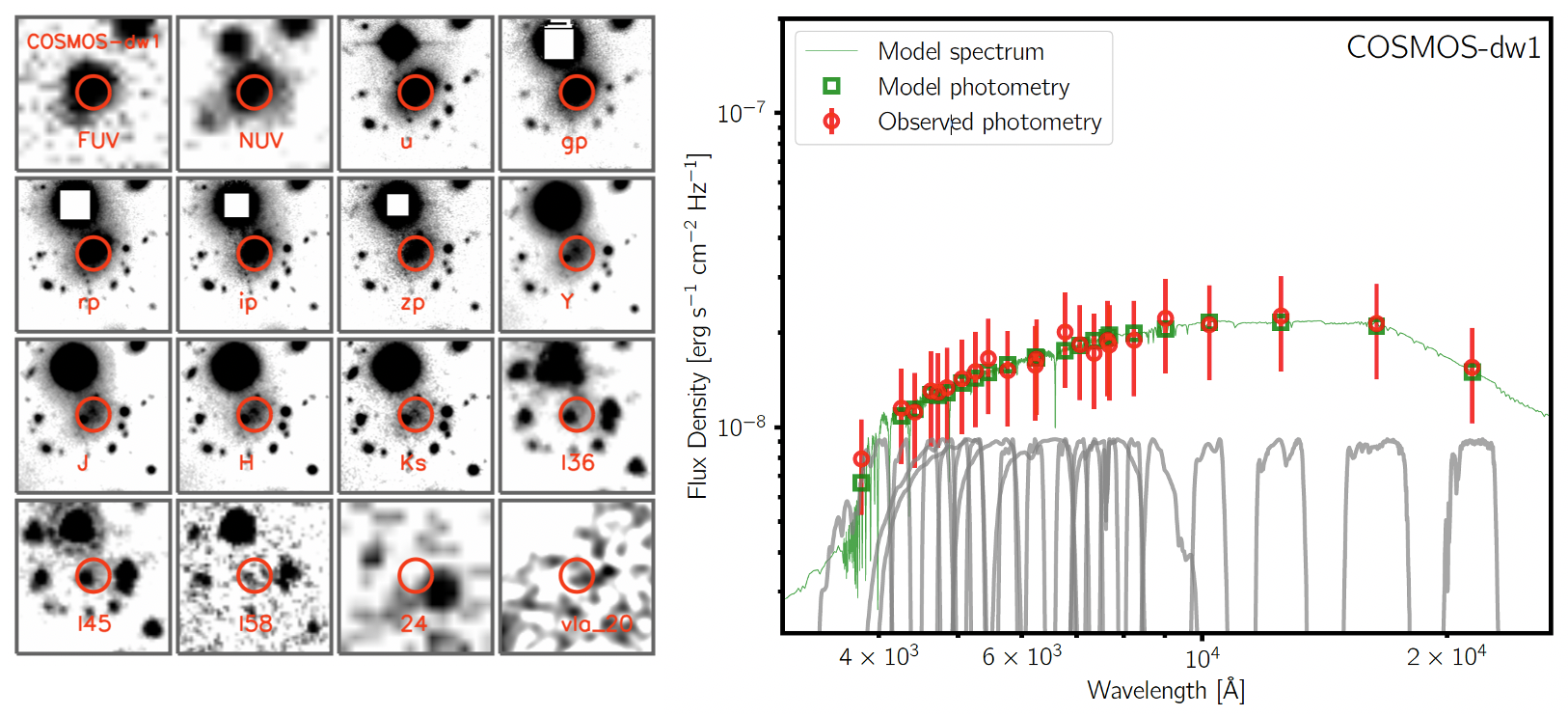}
\caption{ {\bf Left:}The examples of multiwavelength science images of COSMOS-dw1. The size of each stamp is $30\arcsec \times 30\arcsec$. The red circle in each stamp is the target. {\bf Right:} The  SED fitting of COSMOS-dw1. The green curve presents the best-fit model from {\it Prospector}. The grey curves are the filters from optical to NIR. The red points are the observed photometry for COSMOS-dw1, and the green points are the model photometry.    } 
\label{fig:fig2}
\end{center}
\end{figure*}

\begin{figure*}
\begin{center}
\includegraphics[trim=0mm 0mm 0mm 0mm,clip,height=0.42\textwidth]{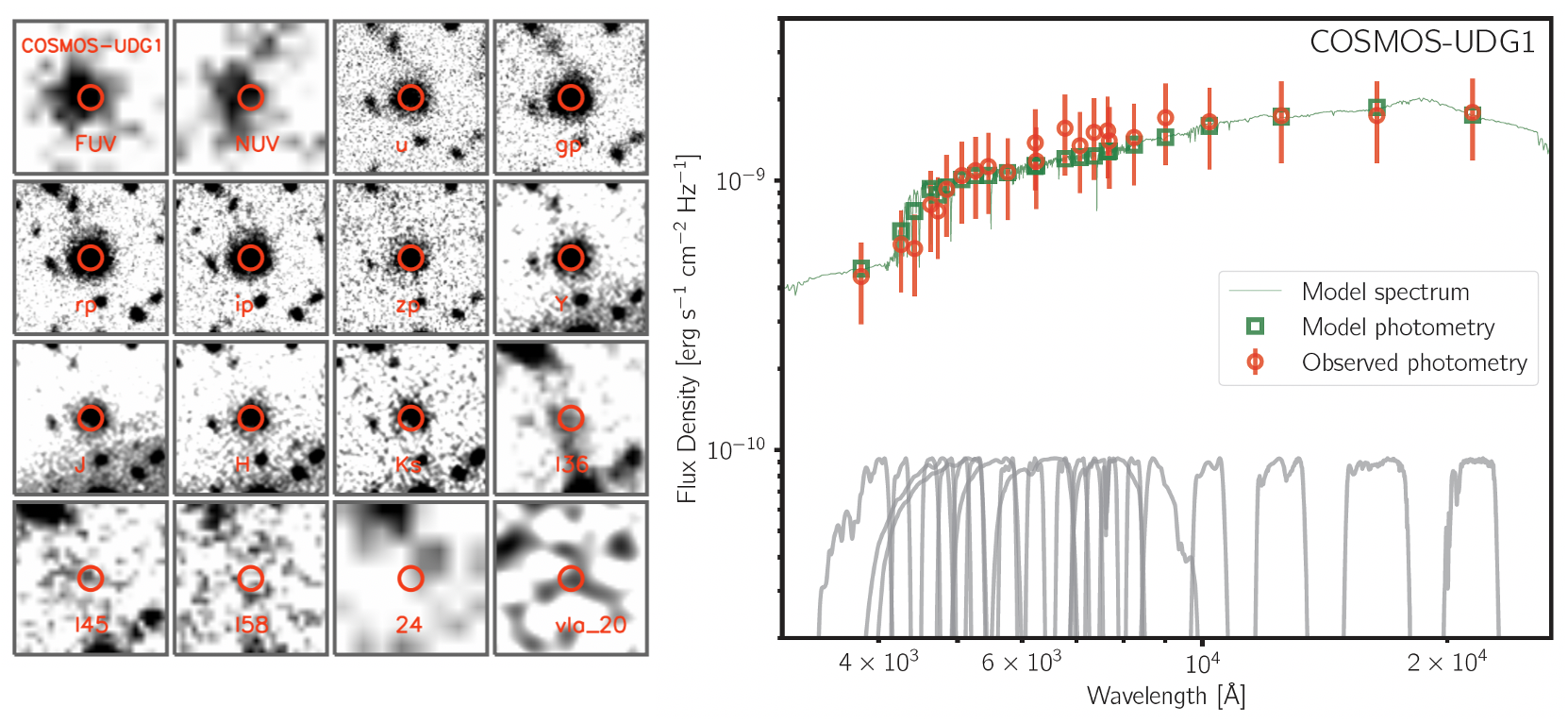}
\caption{{\bf Left:}The examples of multiwavelength science images of COSMOS-UDG1. The size of each stamp is $15\arcsec \times 15\arcsec$. The red circle in each stamp is the target. {\bf Right:} The  SED fitting of COSMOS-UDG1. The green curve presents the best-fit model from {\it Prospector}. The grey curves are the filters from optical to NIR. The red points are the observed photometry for COSMOS-UDG1, and the green points are the model photometry.  }
\label{fig:fig3}
\end{center}
\end{figure*}

\begin{figure*}
\begin{center}
\includegraphics[trim=0mm 0mm 0mm 0mm,clip,height=0.42\textwidth]{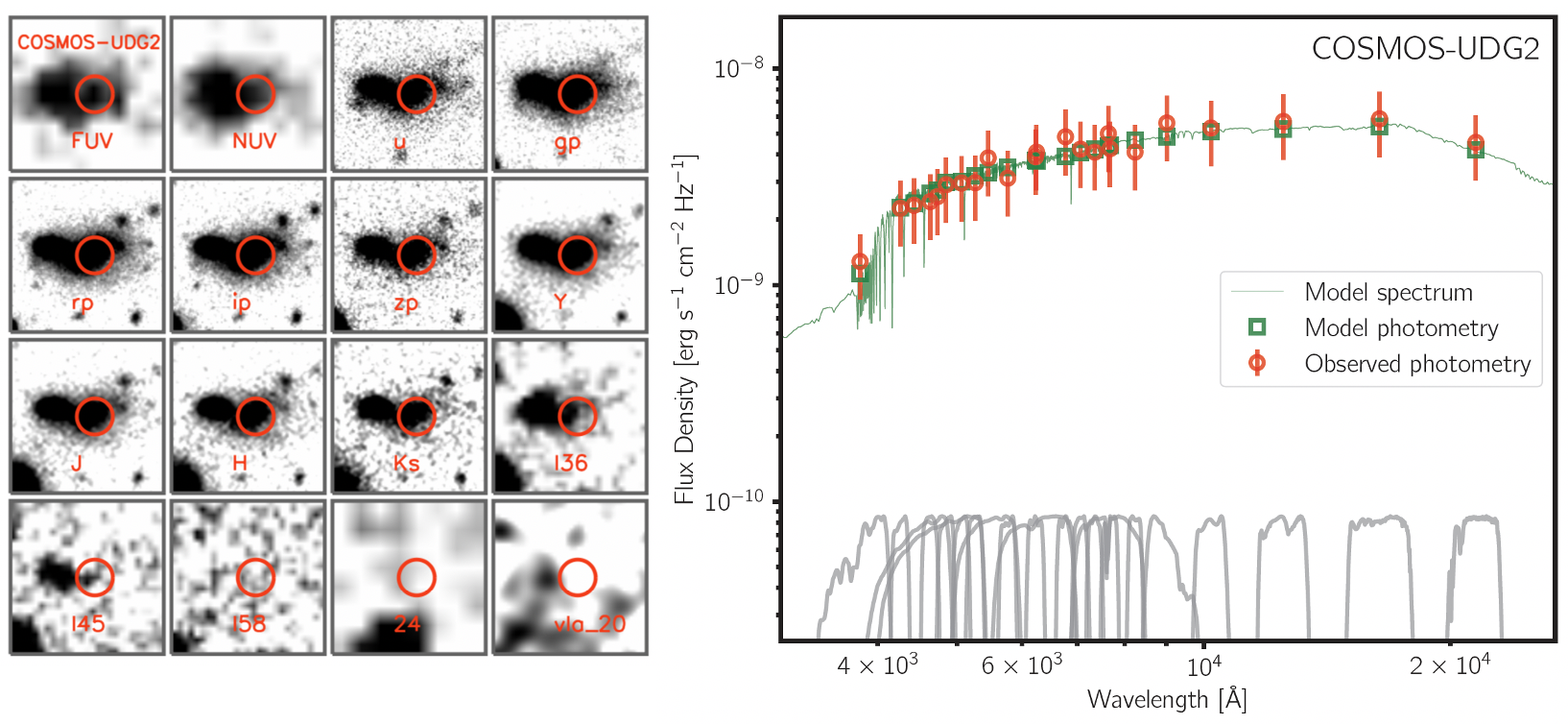}
\caption{{\bf Left:}The examples of multiwavelength science images of COSMOS-UDG2. The size of each stamp is $15\arcsec \times 15\arcsec$. The red circle in each stamp is the target. {\bf Right:} The  SED fitting of COSMOS-UDG2. The green curve presents the best-fit model from {\it Prospector}. The grey curves are the filters from optical to NIR. The red points are the observed photometry for COSMOS-UDG2, and the green points are the model photometry. } 
\label{fig:fig4}
\end{center}
\end{figure*}

The publicly-available multi-band mosaic science images of COSMOS are used to examine the broad properties of the selected UDGs, including Far-UV (FUV) and Near-UV (NUV) images  from the Galaxy Evolution Explorer (GALEX) \citep{Zamojski2007},  $u$ and $i$-band images  obtained with Canada-France-Hawaii Telescope (CFHT) \citep{McCracken2010}, Subaru $gp, rp, ip, zp$ and 12 intermediate-band optical images \citep{Taniguchi2007}, deep $Y, J, H,$ and $K_{\rm s}$-band images  from UltraVISTA \citep{McCracken2012}, IRAC 3.6\,$\mu$m, 4.5\,$\mu$m, 5.8\,$\mu$m, 8\,$\mu$m data from the COSMOS Spitzer survey \citep{Sanders2007}, and 20\,cm data obtained with Very Large Array (VLA) \citep{Schinnerer2007}.  More details about these archive data in COSMOS are summarized in \cite{Laigle2016}.  The left panels of
Figure~\ref{fig:fig2}, Figure~\ref{fig:fig3} and Figure~\ref{fig:fig4} show the representative multiwavelength images of the three galaxies. 

\section{Photometry and Analysis} \label{sec:analysis}

\subsection{Aperture-Matched Photometry} \label{sec:phot}

We construct aperture-matched SEDs from the FUV to the NIR for our three UDG targets.  The three targets are very extended and removal of the blending fluxes from nearby sources is key to measuring their fluxes. The left panels in Figure~\ref{fig:fig2}, Figure~\ref{fig:fig3} and Figure~\ref{fig:fig4} are the examples of mosaics in these three galaxies. Below we describe  our processes to derive the aperture-matched photometry from the multi-band imaging data.

For each of our three targets, we cut stamp images of 30$\arcsec\times30\arcsec$ centered at the target from mosaic science images for further analysis.  We  extract the empirical Point Spread Functions (PSF) from the mosaic science images using the software PSF Extractor \citep[PSFEx, version~3.9.1,][]{Bertin2011}.  SExtractor \citep{Bertin1996} is used to detect sources and  extract their photometric and geometric parameters, including coordinates, magnitude, effective radius $r_{\rm e}$, axis ratio (b/a) and position angle (PA). The detection configuration is optimized for individual stamp images. All sources in one stamp image are simultaneously fitted with 2-D S{\'e}rsic models using GALFIT \citep{Peng2002, Peng2010}.  The best-fit S{\'e}rsic models of detected sources are subtracted from the stamp image and the central target is left. Doing so we obtained clean images of the target.  These clean images are used to match PSFs between different bands and derive aperture-matched photometry.  The total magnitude in $ip$ is estimated using the growth curve derived from the $ip$-band clean image.

We notice that  the total magnitude in {\it HST} $ I_{814}$ is systematically lower than that in $ip$ for all three UDGs. The discrepancy still exists even we measure the total magnitude from the $ I_{814}$ image degraded from a pixel scale of $0\farcs03$ to $0\farcs15$ (the pixel scale of $ip$). We point out that this discrepancy is caused by the background subtraction in $ I_{814}$ data reduction, for which the box size chosen for background estimate is preferentially optimized for faint and compact sources, but too small for extended UDGs. This leads to an oversubtraction of the outskirts of UDGs and therefore the total magnitude to be lower.  The magnitude discrepancy is at a level of $\sim 0.2$\,mag and have marginal effects on the estimate of their geometric parameters. We adopt the parameters from the GALFIT best-fit S{\'e}rsic models of $ I_{814}$ images and list total magnitude of $ V_{606}$, $I_{814}$, $r_{\rm e}$, S{\'e}rsic index $n$, $b/a$ of our three galaxies in Table~\ref{tab:tab1}. 

The Subaru optical images we used are already PSF-matched.  We use a fixed aperture of radius two times the $ip$-band effective radius $r_{\rm e,ip}$ to derive the aperture-matched fluxes from the clean images in the  $gp, rp, ip,$ and $ zp$ and other 12 intermediate bands.  For other images, we match the images of a given band and the $ip$-band to the identical spatial resolution, and then derive aperture-matched flux ratio between the two bands. In practice, we convolve one image with the PSF of the other image and vice versa to match two images to the same spatial resolution.  This method works well for our targets because they are extended and relatively bright.  Aperture photometry on the PSF-matched images with the same circular aperture (i.e., radius=2$\times r_{\rm e,I_{814}}$) gives aperture-matched flux ratio of the two bands. We derived flux ratios of CFHT $u$ and $i$, UltraVISTA $Y, J, H,$ and $K_{\rm s}$ to Subaru $ip$, flux ratios of FUV/NUV to $u$, and flux ratios of  IRAC 3.6\,$\mu$m/4.5\,$\mu$m to $K_{\rm s}$.  These flux ratios based on aperture-matched photometry describe the observed SED over these bands.  The observed SED is then normalized to the total magnitude of $ip$. Taken together, we obtained the FUV-to-NIR SEDs for our three UDG targets, as shown in the right panels of Figure~\ref{fig:fig2}~to~\ref{fig:fig4}.

We also examine the dust emission of our three galaxies using the deepest 850\,$\mu$m map obtained with  SCUBA-2 on board James Clerk Maxwell Telescope (JCMT) through the S2COSMOS survey  \citep{Casey2013, Geach2017, Michalowski2017}, finding no detection at a  level of 3\,$\sigma=1.2$\,mJy.  These suggest a very low rate of obscured star formation among our sample objects. 

\subsection{Modeling of the observed SEDs} \label{sec:seds}

Using the software Easy and Accurate Redshifts from Yale (EAZY) \citep{Brammer2008}, we can estimate photometric redshift (photo-$z$)  from the multiwavelength photometric data.  
The default library of galaxy templates in EAZY is adopted. The input parameters (e.g., templates, input file, output files and redshift grid) are set in the configuration file. The redshift range we have set is from 0 to 2, incremented by 0.001 each step.  Therefore, the photo-$z$  for COSMOS-dw1, COSMOS-UDG1 and COSMOS-UDG2 are $0.010$, $0.130$ and $0.049$, which are fully consistent with the photo-$z$ provided by COSMOS2015 catalog. However, considering that COSMOS-dw1 already has a spectral redshift (spec-$z=$0.0041), we use the spec-$z$ in following SED fitting. The results are listed in Table~\ref{tab:tab1}. 

We constrain the star formation histories (SFHs) of these three galaxies using the SED fitting technique. The fitting is performed using {\it Prospector} \citep{Leja2017, Leja2019}, which uses the Flexible Stellar Population Synthesis (FSPS) package with fully Bayesian Bayesian Markov chain Monte Carlo (MCMC) code \citep{Conroy2009}. We use the default Stellar Population Synthesis (SPS) parameters in FSPS. For the SED modeling, we adopt the \cite{Chabrier2003} IMF and the \cite{Calzetti2000} dust attenuation law. The delayed exponentially declining SFH (SFR$\sim t \times \rm exp(-t/\tau)$) is used, where $t$ is time from the formation. The redshift is fixed by spec-$z$ or photo-$z$. 

The best-fit models of these three galaxies with {\it Prospector} are shown in the right panels of the Figure~\ref{fig:fig2}, Figure~\ref{fig:fig3} and Figure~\ref{fig:fig4}.  For COSMOS-dw1, we obtain stellar mass $M_{\ast}= 5.6_{-2.7}^{+2.5}\times10^{6}$\,M$_{\odot}$, the same as calculated in \cite{Polzin2021}. The estimated stellar metallicity, star formation rate (SFR) and stellar age of COSMOS-dw1 is log(Z/Z$_{\odot}$) $=-1.47_{-0.39}^{+0.63}$, 0.001\,$M_{\odot}$\,yr$^{-1}$ and $4.10_{-2.5}^{+5.24}$\,Gyr, respectively.  \cite{Polzin2021} claim that COSMOS-dw1 is an isolated quenched low-mass galaxies with strong Balmer absorption lines in the local group, but the specific stellar age and metallicity are not given. Our results support that COSMOS-dw1 has very little  star formation at present.  For COSMOS-UDG1, we obtain stellar mass $M_{\ast}= 2.5_{-3.4}^{+1.9}\times10^{8}$\,M$_{\odot}$, stellar metallicity log(Z/Z$_{\odot}$) $=-0.74_{-0.86}^{+0.54}$,  stellar age $1.70_{-1.43}^{+4.41}$\,Gyr.  The parameter of stellar mass, stellar metallicity and stellar age in COSMOS-UDG2 is  $M_{\ast}= 2.2_{-1.1}^{+0.9}\times10^{8}$\,M$_{\odot}$, log(Z/Z$_{\odot}$) $=-1.44_{-0.40}^{+0.65}$ and $5.58_{-3.12}^{+5.10}$\,Gyr, respectively. We find that these three galaxies exhibit diverse properties. COSMOS-UDG1 and COSMOS-UDG2 have similar stellar masses, but COSMOS-UDG1 is younger and more metal-rich than COSMOS-UDG2 and COSMOS-dw1. COSMOS-UDG2 and COSMOS-dw1 have similar stellar metallicities, but COSMOS-UDG2 is older than COSMOS-dw1.

\begin{figure*}
 \begin{center}
  \includegraphics[height=0.7\textwidth]{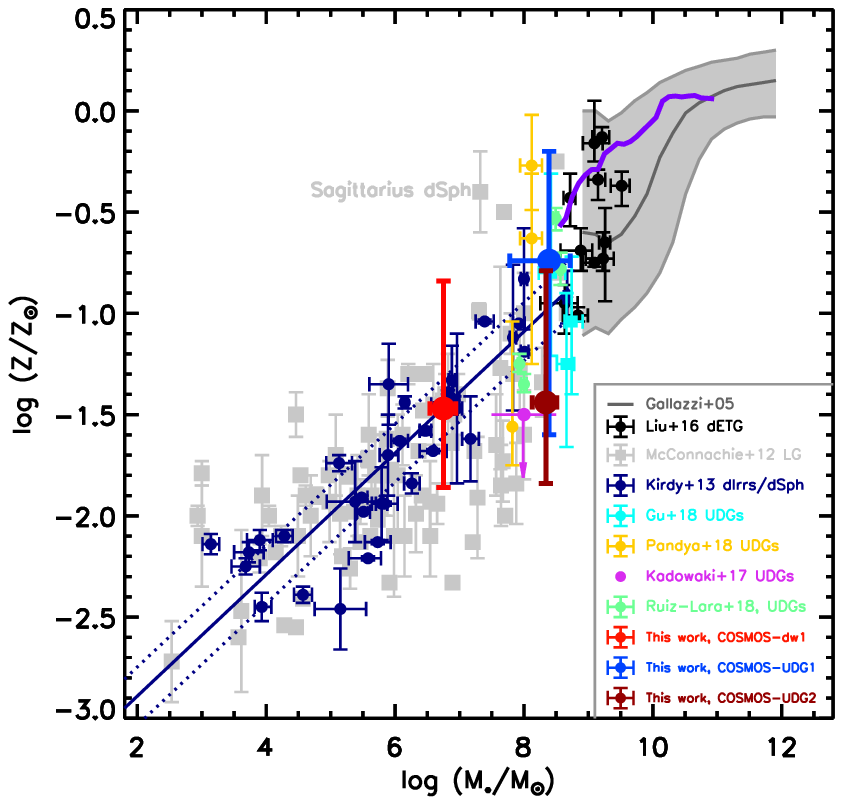}
  \caption{The stellar mass-metallicity relation. The black solid points are the early-type galaxies (ETGs) in Virgo \citep{Liu2016}, dwarf galaxies in and around the Local group are shown in gray solid squares \citep{McConnachie2012}, and local group dIrrs/dSphs from \cite{Kirby2013} are shown in dark blue solid points, the blue solid line shows the least-squares line, and the dotted lines are the rms about the best fit. MZR relation could extend to the massive galaxies \citep{Gallazzi2005}, as shown the gray solid curves. The purple solid line is the MZR relation of star-forming galaxies with stellar masses ranging between $10^{8.5}$ and $10^{11}$\,M$_{\odot}$ at $z=0.027-0.25$ \citep{Zahid2017}. Other UDGs from literatures are presented for comparisons \citep[e.g.,][]{Kadowaki2017, Gu2018, Pandya2018, Ruiz-Lara2018}. The red, blue and brown solid points stands for the COSMOS-dw1, COSMOS-UDG1 and COSMOS-UDG2, respectively.}
 \label{fig:fig5}
 \end{center}
\end{figure*}

In Figure~\ref{fig:fig5}, we show the stellar mass–metallicity relation (MZR) for these three COSMOS galaxies, compared to the relation found for the dwarfs \citep{Kirby2013} and giant galaxies in the local Universe \citep{Gallazzi2005}. The MZR with large scatter appears to be continuous from low to high masses. Despite the large uncertainties of metallicity and stellar mass, COSMOS-dw1 obeys the MZR at the low mass end, COSMOS-UDG1 and COSMOS-UDG2 follow the MZR defined by normal dwarf galaxies.  This suggests that stellar mass plays an important role in determining stellar metallicities, regardless of the size of a galaxy.

\section{Discussion and Summary} \label{sec:discussion}

We present the physical properties of three nearby diffuse galaxies identified in the central region of the COSMOS field, which is covered by the 3D-HST/CANDELS survey. The primary uncertainty in our analysis is the distances of the three diffuse galaxies. The photometric redshifts derived from the broadband SEDs affirm that these objects are nearby, with redshifts $z<0.15$. Nevertheless, given uncertainties associated with photo-$z$, we cannot rule out the possibility that they might be located closer (e.g., in the Local Group), especially for COSMOS-dw1. COSMOS-dw1 has been confirmed by optical spectroscopy and radio observations to have a redshift of 0.004 \citep{Polzin2021,Pan2024}. For  COSMOS-UDG1 and  COSMOS-UDG2, we determine their photo-$z$s to be $0.130$ and $0.049$, respectively, using multiwavlength data. The physical properties of these three galaxies appear to be strikingly different. 

Obtaining accurate distance estimates for the ultra-faint and diffuse objects in local Universe is critical to derive correct galaxy physical properties. \cite{Polzin2021} applied the surface brightness fluctuation (SBF) method to COSMOS-dw1 and measured a distance of $22\pm3$\,Mpc, which aligns with its radial velocity of $1222\pm64$\,km\,$s^{-1}$. However, recent work by \cite{Foster2024} used the SBF method to derive distance estimates for the 20 nearby dwarf galaxies detected in the COSMOS field, with COSMOS-dw1 being one  of them. The SBF distance is estimated to be $56.3_{-6.7}^{+10.4}$\,Mpc \citep{Foster2024}, which is three times higher than estimated provided by \cite{Polzin2021}.  
Although \cite{Foster2024} can recover a similar result ($23\pm5$\,Mpc) as \cite{Polzin2021} by modifying certain methodologies, but SBF distance estimates for the rest galaxies will be severely underestimated ($\sim10-30$\,Mpc) after the same modified method is applied to the whole sample. If we adopt the distance of $56.3_{-6.7}^{+10.4}$\,Mpc, the estimated stellar mass $M_{\ast}$ of COSMOS-dw1 would be an order of magnitude higher than previously estimated, and its effective radius would be three times larger than estimated in the Table~\ref{tab:tab1}.  Therefore, we emphasize that there are certain discrepancies in the distances derived using the SBF method.

COSMOS-dw1 exhibits a blue color with $V_{606}-I_{814}=0.11\pm0.05$, and has been detected in radio observations to possess H$\small \rm I$ gas. The  H$\small \rm I$ mass is $M_{\rm HI}=4.90\pm0.90\times10^{6}$\,M$_{\odot}$, with the line width $W_{50}$ is 18.2\,km\,s$^{-1}$, as reported by \citep{Pan2024}. The gas fraction $M_{\rm HI}/M_{\ast}$ is $0.87\pm0.43$, indicating that this galaxy is not gas-poor and still retains a significant amount of atomic gas despite exhibiting quiescent optical spectra \citep[top panel of figure~2]{Polzin2021}. Given the isolation, COSMOS-dw1 is unlikely to have undergone strong environmental effects \citep{Polzin2021}.  Furthermore, the stellar age of COSMOS-dw1 is estimated to be 4.1\,Gyr, suggesting this galaxy formed at $z\sim0.38$.

The dynamical mass $M_{\rm dyn}$ is estimated using the formula $M_{\rm dyn}=3.5\times10^{5} r_{\rm e} W_{50}^{2}$\,$M_{\odot}$ from \cite{Spekkens2018}, where $r_{\rm e}$ is effective radius kpc and $W_{50}$ is the line width in units of  km\,s$^{-1}$.  COSMOS-dw1 has $W_{50}=18.2\,\rm km\,s^{-1}$ \citep{Pan2024}. Therefore, we calculate the dynamical mass of COSMOS-dw1 to be $3.36\pm0.12\times10^{7}\,M_{\odot}$. Additionally, we estimate the baryonic mass of COSMOS-dw1 as $M_{\rm bar}=1.33M_{\rm HI}+M_{\ast}=1.21\pm0.29\times10^{7}\,M_{\odot}$ \citep{Pina2019}. Assuming the cosmological baryon fraction is 0.16, we derive the virial mass of the dark matter halo to be $7.59\pm1.78\times10^{7}\,M_{\odot}$, which is $2-3$ times higher than the calculated dynamical mass.

The MZR of galaxies offers profound insights into their star formation and chemical enrichment histories. The relatively low scatter in this relation, particularly at the low-mass end \citep[e.g.,][]{Kirby2013}, poses a challenge to explain. This relationship is intricately tied to the complex dynamics involving reionization, star formation, gas inflow, outflow, and recycling processes \citep[e.g.,][]{Ma2016}. 

Recently, numerous researches have unveiled the stellar population properties of some UDGs through optical spectra and multiwavelength photometric data in different environments. These studies have demonstrated the diverse stellar populations of UDGs across different environments. Specifically, UDGs in clusters (e.g., Coma and Virgo) identified by optical spectroscopy are intermediate-to-old age ($>6-10$\,Gyr) and metal-poor \citep[e.g.,][]{Kadowaki2017,Ferre-Mateu2018, Gu2018,Ruiz-Lara2018,Villaume2022, Buzzo2022, Ferre-Mateu2023, Gannon2024, Buzzo2024a,Buzzo2024b}.   In contrast, some star-forming UDGs in low-density environments are significantly more metal-rich and younger ($\sim<5$\,Gyr) compared to their quiescent counterparts \citep[e.g.,][]{Martinez-Delgado2016, Trujillo2017, Rong2020}.  Using multiwavelength photometric data, several studies have further revealed that UDGs found in clusters are older than those in the field \citep{Pandya2018, Buzzo2022}. Additionally, some field UDGs showcase stellar populations of intermediate age on average ($\sim7$\,Gyr), with some being metal-poor and others metal-rich \citep{Barbosa2020}. 

We examine the environments around the three galaxies, and find that COSMOS-dw1, COSMOS-UDG1 and COSMOS-UDG2 do not have obvious luminous companions, suggesting that all three galaxies reside in the low-density environments. In comparison to UDGs in galaxy clusters \citep[e.g.,][]{Kadowaki2017,Gu2018,Ruiz-Lara2018, Buzzo2022}, COSMOS-UDG1 shows younger age and higher metallicity, whereas COSMOS-UDG2 is younger but metal-poor. These observations imply that the relative young ages of COSMOS-UDG1 and COSMOS-UDG2 may be associated with their low-density environment \citep{Martinez-Delgado2016, Trujillo2017, Pandya2018}.  A possible explanation is that COSMOS-UDG1 have relatively massive halos, more metals can be locked, and finally reproduce the metal-rich galaxies. Besides, the non-universal initial mass function (IMF) may provide the constrains \citep{Ferre-Mateu2013}. Interestingly, the gray squares in Figure~\ref{fig:fig5} show the Sagittarius (Sgr) dwarf spheroidal (dSph) galaxy, a satellite galaxy in the Milky way, exhibits a relatively high metallicity ([Fe$/$H]$\sim0.4$) despite stellar mass is comparable to those of UDGs \citep{Chou2007, McConnachie2012}, which is consistent with the results of COSMOS-UDG1. 

COSMOS-UDG1 and COSMOS-UDG2 belong to the dwarf galaxies with large size, and their diffuse nature potentially may be governed by internal mechanisms. UDGs are the extended dwarf galaxies with high spin angular momentum \citep{Amorisco2016, Rong2017}, and strong feedback from supernova or massive stars driven gas outflow, dark matter halo and stellar disks expansion, and reproduce low luminosity and extended galaxies \citep{DiCintio2017, Chan2018}. 
Furthermore, some UDGs may be tidal disturbed dwarf galaxies and some present tidal feature associated with galaxy mergers \citep{Merritt2016, Greco2018a}. From the deep multiwavelength imaging, the three COSMOS galaxies we identified appear to be no tidal structures.  The multiwavelength photometric data can help constrain the properties of these three galaxies, and we look forward to spatially resolving these diffuse galaxies in subsequent work to understand their formation mechanisms.

We summarize our results as follows:

\begin{itemize}
\item[(1)] We conducted a search for a low-mass LSB galaxies (COSMOS-dw1) and two new UDGs (COSMOS-UDG1 and COSMOS-UDG2) within the central region of the COSMOS field, and examine their properties using the existing multiwavelength data. We present their UV-to-IR SEDs built through our careful PSF- and aperture-matched photometry. The spec-$z$ or photo-$z$ in COSMOS-dw1, COSMOS-UDG1 and COSMOS-UDG2 is 0.004, 0.130 and 0.049, respectively.

\item[(2)]  SED fitting reveals that these three galaxies exhibit different physical properties. COSMOS-dw1 is a quenched low-mass galaxy with a stellar mass of $5.6_{-2.7}^{+2.5}\times10^{6}$\,M$_{\odot}$. The stellar age and metallicity log(Z/Z$_{\odot}$) of COSMOS-dw1 is $4.10_{-2.5}^{+5.24}$\,Gyr and  $-1.47_{-0.39}^{+0.63}$, respectively.  COSMOS-UDG1 and COSMOS-UDG2 have similar stellar masses ($\sim10^{8}\,M_{\odot}$), yet COSMOS-UDG1 is younger and more metal-rich than COSMOS-UDG2 and COSMOS-dw1. COSMOS-UDG2 and COSMOS-dw1 exhibit comparable stellar metallicities, but COSMOS-UDG2 is older than COSMOS-dw1.  When compared  to cluster UDGs \citep[e.g.,][]{Kadowaki2017,Gu2018,Ruiz-Lara2018, Buzzo2022}, COSMOS-UDG1 shows younger age and higher metallicity, whereas COSMOS-UDG2 is younger and metal-poor. This hints that the relatively young ages of COSMOS-UDG1 and COSMOS-UDG2 may be associated with their low-density environment.

\item[(3)] Interestingly, COSMOS-dw1 contains atomic gas with an H$\small \rm I$ mass of $4.90\pm0.90\times10^{6}$\,M$_{\odot}$, and gas fraction ($M_{\rm HI}/M_{\ast}$) is $0.87\pm0.43$. This indicates that this galaxy may be in the initial stage of quenching. The estimated dynamical mass is about $3.4\times10^{7}\,M_{\odot}$, implying that COSMOS-dw1 is dominated by dark matter ($>60\%$).

\item[(4)]  Despite the significant uncertainties in metallicity measurements, COSMOS-dw1 aligns with the MZR at the low mass end, while COSMOS-UDG1 and COSMOS-UDG2 broadly follow the MZR established by typical dwarf galaxies. This suggests that stellar mass may be a crucial factor in determining stellar metallicities. 
\end{itemize}

Taken together, the detection of low-luminosity LSB galaxies and UDGs in the COSMOS field indicate that UDGs can indeed be found in random fields. These extreme LSB galaxies identified  so far are just the tip of the iceberg. In the future, more unknown LSB galaxies and UDGs could be discovered through multiwavelength imaging facilitated by space- (e.g., Euclid, CSST and Roman) \citep[e.g.,][]{Zhan2011,Montes2023,Mellier2024} and ground-based (e.g., LSST and WFST) telescopes with a wide field-of-view capabilities \citep[e.g.,][]{Robertson2017,Shi2018,Martin2022,Breivik2022,Wang2023}. By combining these observations with spectroscopic analysis, the real natures of UDGs and LSB structures can be fully unveiled. 

\clearpage
\appendix
\section{APPENDIX: The Parameter measurements of three UDGs}

\begin{figure*}
 \begin{center}
    \includegraphics[trim=0mm 6mm 0mm 0mm,clip,height=0.67\textwidth]{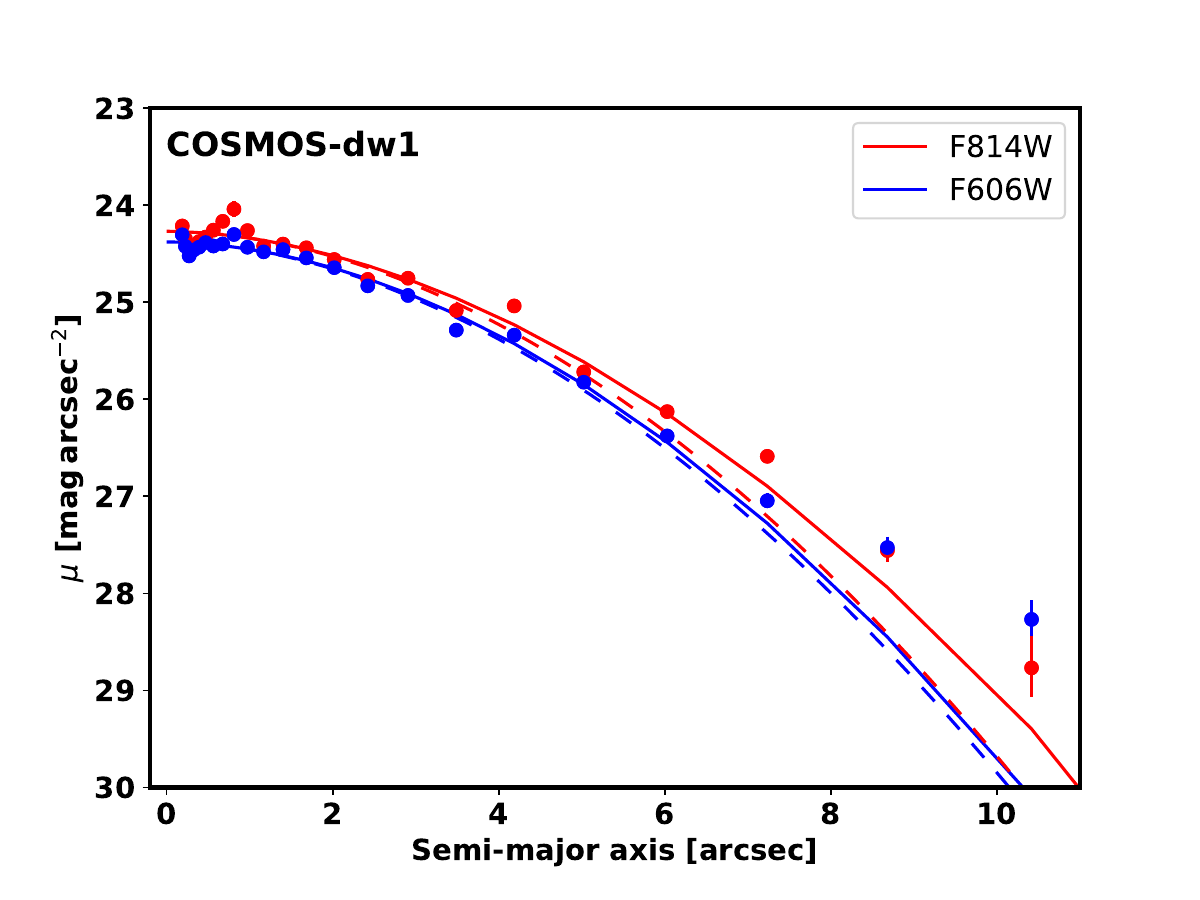}
    \includegraphics[height=0.56\textwidth]{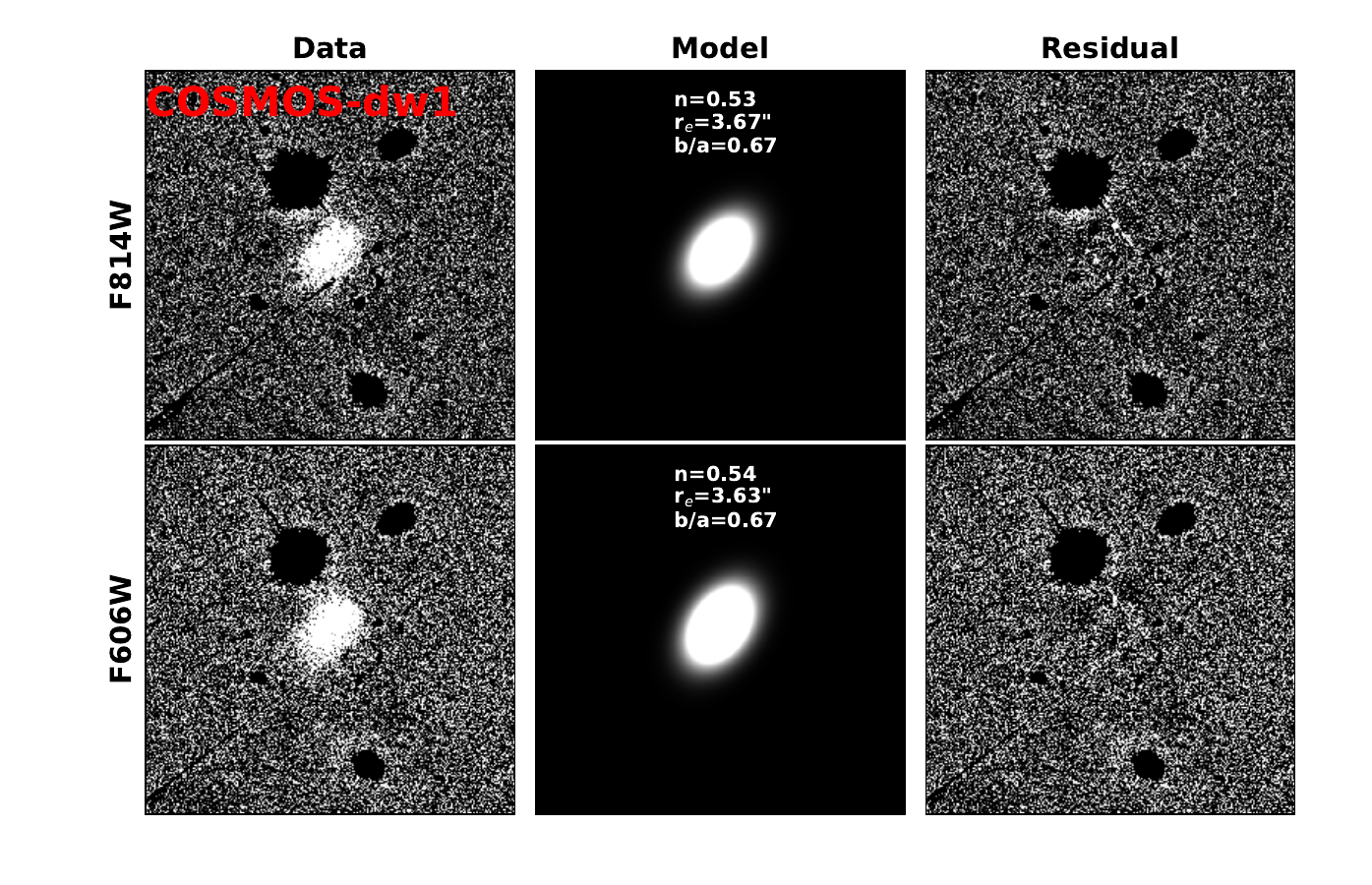}
  \caption{The 1D surface brightness profile (Upper panel) and 2D S{\'e}rsic fitting (Bottom panel) of COSMOS-dw1. The size of each stamp is $48\arcsec \times 48\arcsec$.}
  \label{fig:figa1}
 \end{center}
\end{figure*}

\begin{figure*}
 \begin{center}
    \includegraphics[trim=0mm 6mm 0mm 0mm,clip,height=0.67\textwidth]{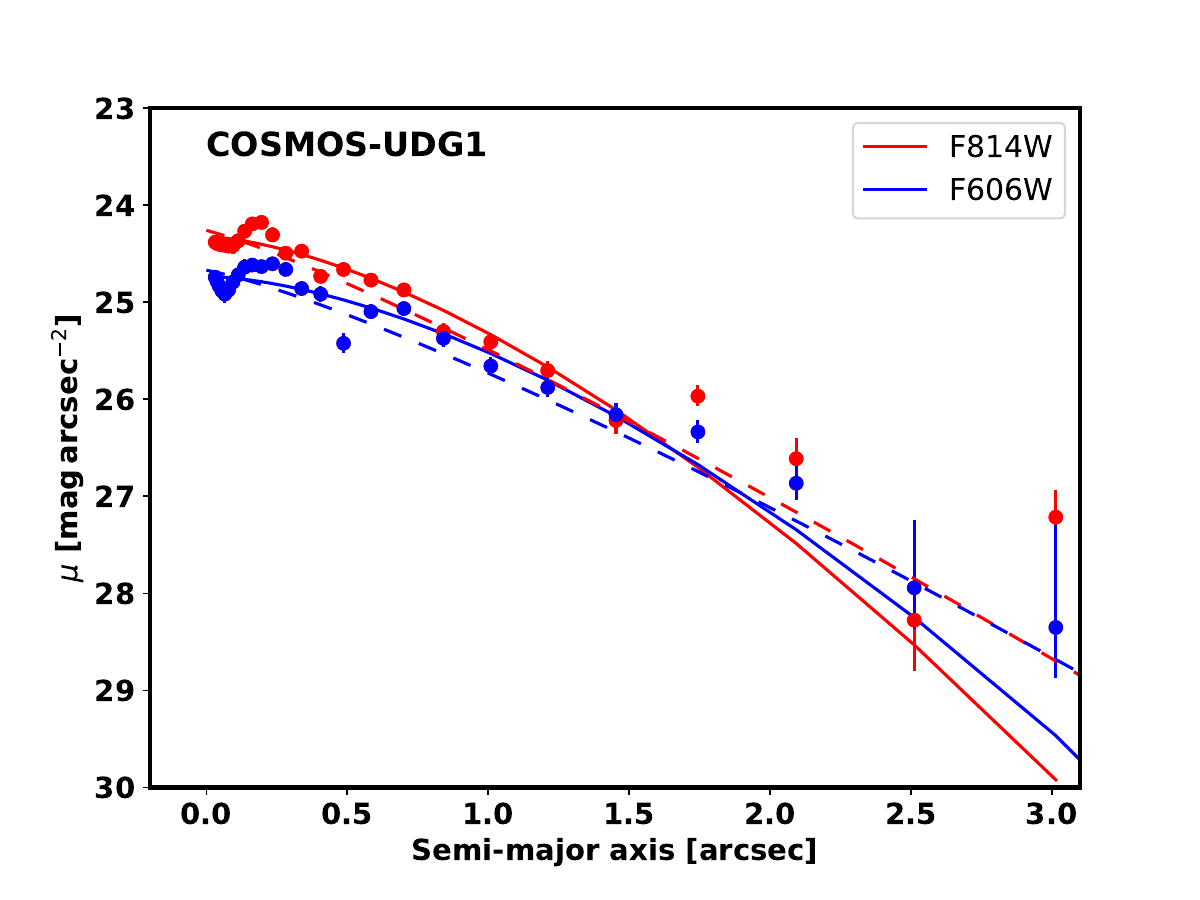}
    \includegraphics[height=0.56\textwidth]{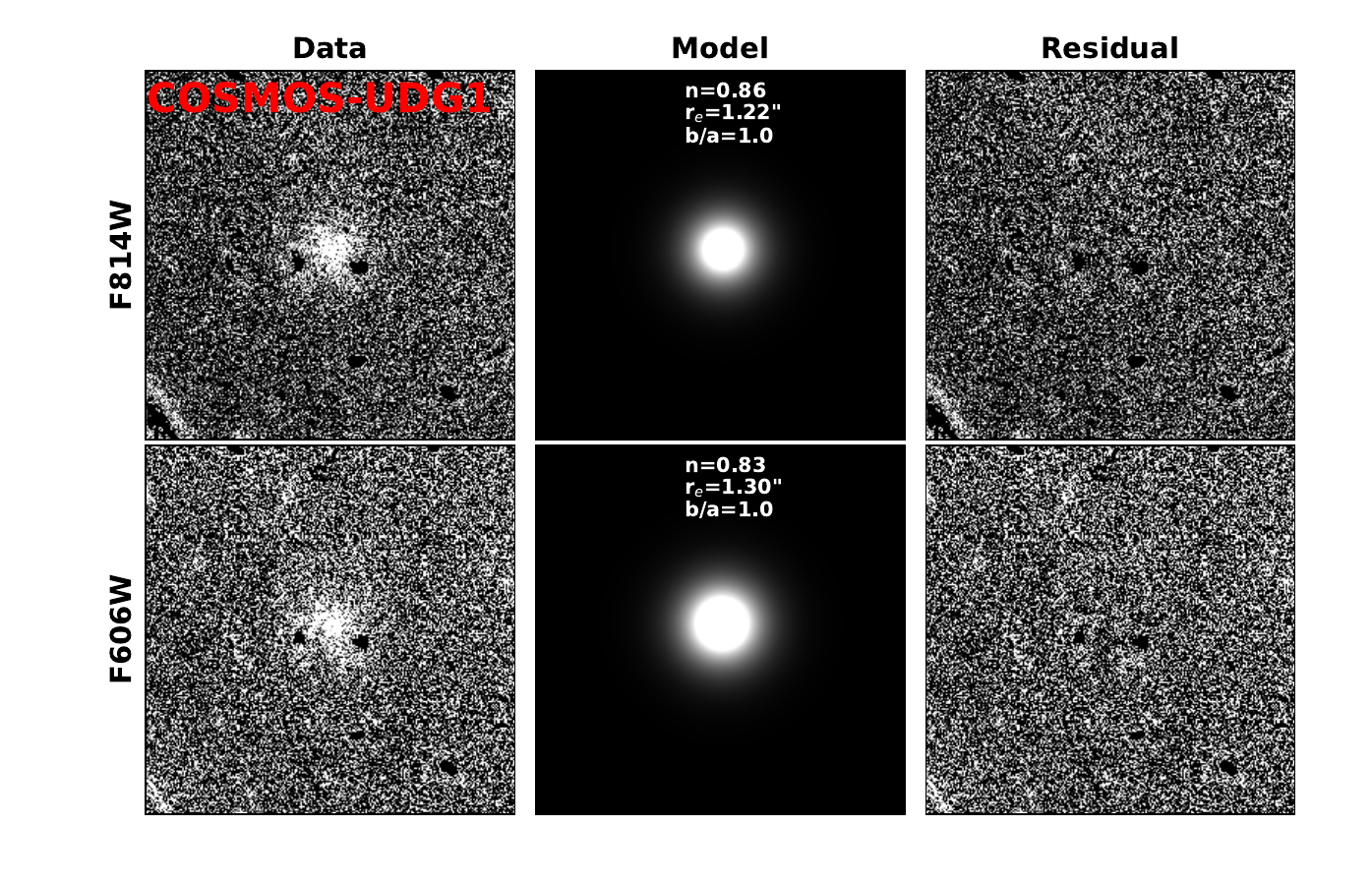}
  \caption{The 1D surface brightness profile (Upper panel) and 2D  S{\'e}rsic fitting (Bottom panel) of COSMOS-UDG1. The size of each stamp is $15\arcsec \times 15\arcsec$.}
  \label{fig:figa2}
 \end{center}
\end{figure*}

\begin{figure*}
 \begin{center}
    \includegraphics[trim=0mm 6mm 0mm 0mm,clip,height=0.67\textwidth]{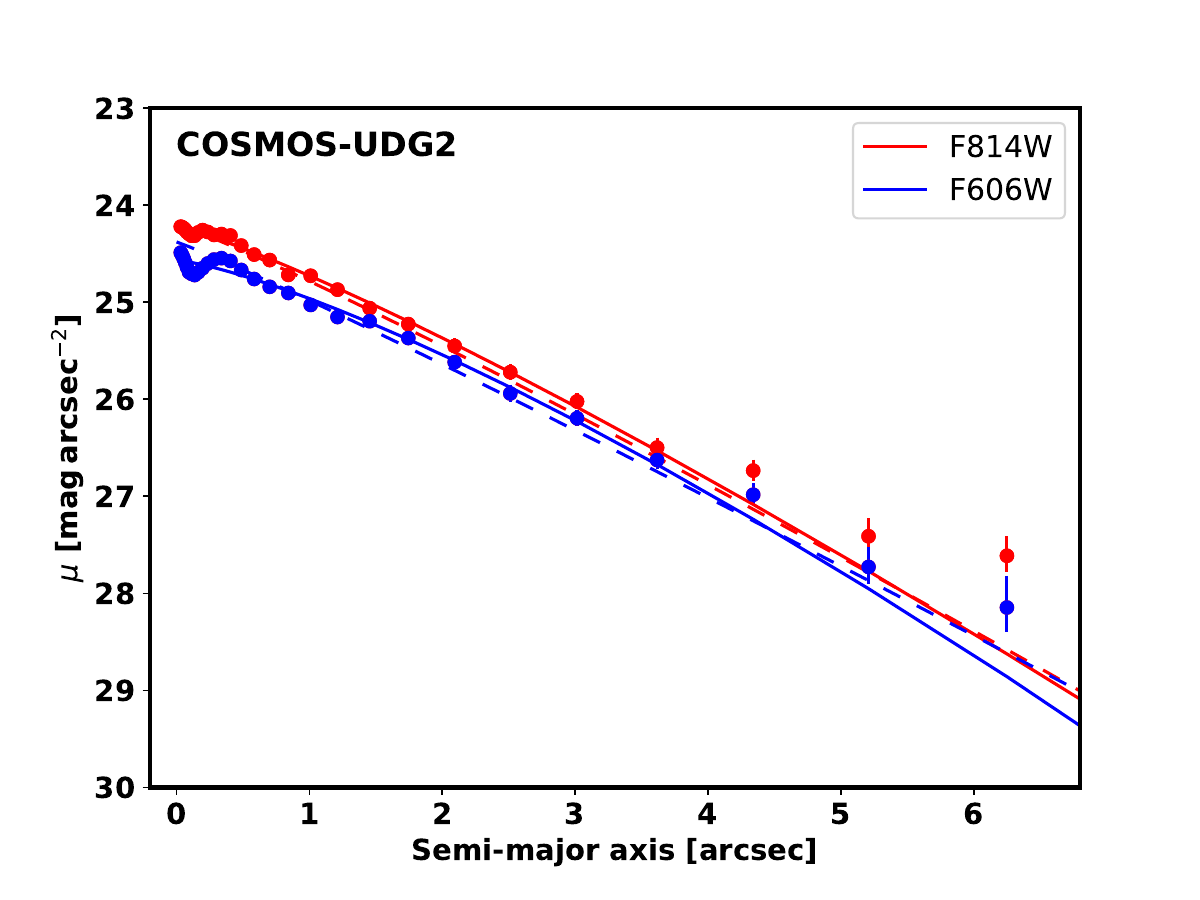}
    \includegraphics[height=0.56\textwidth]{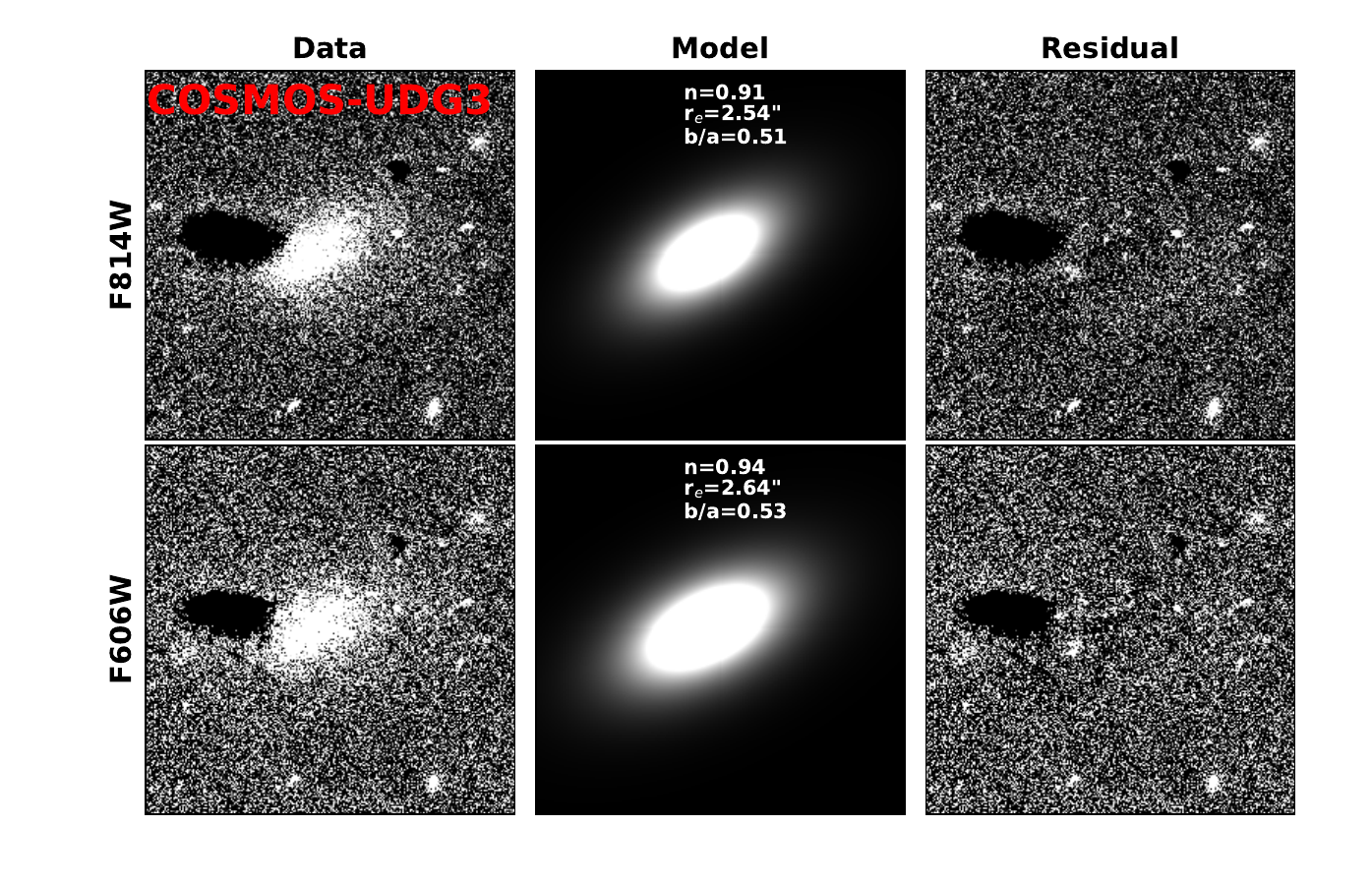}
  \caption{The 1D surface brightness profile (Upper panel) and 2D  S{\'e}rsic fitting (Bottom panel) of COSMOS-UDG2. The size of each stamp is $15\arcsec \times 15\arcsec$.}
  \label{fig:figa3}
 \end{center}
\end{figure*}

Here, we have measured the structure parameters of COSMOS-dw1, COSMOS-UDG1 and COSMOS-UDG2 using both 1D S{\'e}rsic and 2D  S{\'e}rsic fitting methods. The structure properties obtained are summarized in Table~\ref{tab:tab2}.  Figure~\ref{fig:figa1}, Figure~\ref{fig:figa2} and Figure~\ref{fig:figa3} depict the 1D surface brightness profiles and 2D  S{\'e}rsic fitting results for COSMOS-dw1, COSMOS-UDG1 and COSMOS-UDG2, respectively.  In the upper panels of Figures~\ref{fig:figa2} and \ref{fig:figa3}, the blue and red points and curves represent the results from the F606W and F814W filters, respectively. The bottom panels of these figures show, from left to right, the original image, the model, and the residual image for each galaxy. The axial ratios (b/a) of these three diffuse galaxies are greater than 0.5, indicating that they exhibit an ellipsoidal shape. Their S{\'e}rsic indices $\rm n=0.53-0.91$ suggest that they are similar to typical disk galaxies. From the deep imaging, we find that the three COSMOS galaxies do not exhibit tidal features.

\begin{table*}[ht]

\begin{center}

\scriptsize \footnotesize

\caption{ The structure properties of COSMOS-dw1, COSMOS-UDG1 and COSMOS-UDG2. \label{tab:tab2}}

\setlength{\tabcolsep}{0.7mm}{
\begin{tabular}{l|cccc|cccc}

\hline

\hline

&  &  & S{\'e}rsic 1D & & &  S{\'e}rsic 2D & & \\

& & COSMOS-dw1 & COSMOS-UDG1 & COSMOS-UDG2 & COSMOS-dw1 & COSMOS-UDG1 & COSMOS-UDG2 \\

\hline
R.A.\,(J2000.0)  & & ... &... &... & 10:00:30.066 & 10:00:37.857 & 10:00:23.792 \\

Decl.\,(J2000.0)& & ...&... &... & +02:08:58.880 & +02:24:31.990 & +02:22:05.660 \\

$V_{606}$\,(mag) & & 19.95 & 22.94 & 21.30 & 19.95 & 22.94 & 21.30 \\

$I_{814}$\,(mag) &  & 19.72 & 22.82 & 21.11 & 19.72 & 22.82 & 21.11 \\

$V_{606}$-$I_{814}$\,(mag) &  & 0.23 & 0.12 & 0.19 & 0.23 & 0.12 & 0.19 \\

$\mu(\rm V_{606},0)$\,(mag\,arcsec$^{-2}$) &  & 24.38 & 24.73 & 24.57 & 24.33 & 24.66 & 24.38 \\

$\mu(\rm I_{814},0)$\,(mag\,arcsec$^{-2}$) & & 24.27 & 24.33 & 24.22 & 24.14 & 24.35 & 24.16 \\

r$_{\rm e,I_{814}}$\,($\arcsec$) &  & 3.91 & 1.03 & 2.50 & 3.67 & 1.22 & 2.54 \\

S{\'e}rsic index\,(n) &  & 0.55 & 0.64 & 0.85 & 0.53 & 0.86 & 0.91 \\

axis ratio\,(b/a) &  & ... & ... & ... & 0.67 &1.0 & 0.51 \\
\hline

\end{tabular}}
\end{center}
\end{table*}

\section*{Conflict of Interest Statement}

The authors declare that the research was conducted in the absence of any commercial or financial relationships that could be construed as a potential conflict of interest.

\section*{Author Contributions}

D.D Shi and X.Z Zheng designed the study. D.D Shi led data analyses including the deep multiwavelength imaging, and wrote the original manuscript. Z. Pan and Y. Luo made significant contributions and edits to the text.  All authors contributed to the article and approved the submitted version.

\section*{Funding}
This work is supported by Scientific Research Foundation for High-level Talents of Anhui University of Science and Technology (2024yjrc104), the National Science Foundation of China (12303015 and 12173088), and the National Science Foundation of Jiangsu Province (BK20231106). 

\section*{Acknowledgments}
We thank the referees for the valuable and helpful comments and suggestions, which improve our manuscript. 
We acknowledgment support from Anhui University of Science and Technology and China Manned Space Project.


\section*{Data Availability Statement}
The raw data supporting the conclusions of this article will be made available by the authors, without undue reservation.
\bibliographystyle{Frontiers-Harvard} 
\bibliography{ms}



\end{document}